\documentclass[a4paper]{article}
\usepackage{amsmath}
\usepackage{amsfonts}
\usepackage[T1]{fontenc}
\usepackage[english]{babel}
\usepackage[latin1]{inputenc}
\usepackage{ae}
\usepackage{aecompl}
\newtheorem{theorem}{Theorem}
\newtheorem{lemma}[theorem]{Lemma}

\newtheorem{definition}[theorem]{Definition}

\title{Explicit multipole moments of stationary axisymmetric spacetimes}
\author{Thomas B\"ackdahl${}^*$, Magnus Herberthson\thanks{Department of  
Mathematics, Link\"oping University,
SE-581 83 Link\"oping, Sweden.\newline
\hspace*{5mm} e-mail: thbac@mai.liu.se, maher@mai.liu.se}}
\date{}

\begin{document}
\maketitle
\begin{abstract}
In this article we study multipole moments of axisymmetric stationary asymptotically flat
spacetimes. We show how the tensorial recursion of Geroch and Hansen can be reduced
to a recursion of scalar functions. We also demonstrate how a careful choice of conformal factor collects all 
moments into one complex valued function on $\mathbb{R}$, where the moments appear as the derivatives at $0$. 
As an application, we calculate the moments of the Kerr solution. We also discuss 
the freedom in choosing the potential for the moments.

\end{abstract}
\section{Introduction}
The relativistic multipole moments of stationary spacetimes have been defined by Hansen \cite{hansen}.
This definition is an extension of the static case considered by Geroch \cite{geroch}, and apart from
a slightly different setup due to the possible angular momentum, the recursive definitions of the moments 
in \cite{hansen} and \cite{geroch} are the same. 
The Hansen formulation reduces to the Geroch formulation in the static case but with a different potential.
In section \ref{potsec} we conclude that these two potentials indeed give the same multipole moments
in the general axisymmetric static case.
Beig \cite{beigAPA} defined a generalisation of centre of mass so that the expansion of the Hansen moments 
around this 'point' determines the multipole moments uniquely.
Thorne \cite{thorne} gave another definition of multipole moments which is known \cite{gursel} to be
equivalent to the Hansen formulation if the spacetime has non-zero mass.
There are also other definitions of multipole moments \cite{beigAPA}, \cite{clarke}, \cite{burg}, \cite{janis} 
which will not be considered here. See for instance \cite{quevedo} for further details about these moments.

The recursive definition of multipole moments of Geroch and Hansen (\ref{orgrec}) takes place in a conformal 
compactification of the 3-manifold of Killing-tra\-jec\-tor\-ies. The recursion produces a family of totally symmetric
and trace-free tensors, which are to be evaluated at a certain point, and the values will
then provide the moments of the spacetime in question. Even in the case of 
axisymmetric spacetimes, the actual calculations of the tensors in \eqref{orgrec} are non-trivial. 

In \cite{herb}, it was shown how the moments in the axisymmetric static case can be obtained
through a set of recursively defined real valued functions $\{f_n\}_{n=0}^\infty$ on $\mathbb{R}$. The moments 
are then given by the values $\{f_n(0)\}_{n=0}^\infty$.  In this way,
one can easily calculate `any' desired number of moments. By exploring the conformal freedom
of the construction, it was also shown how all moments could be captured in one
real valued function $y$, where the moments appeared as the derivatives of $y$ at $0$.
In this paper we show that a similar scalar recursion can be found in the stationary axisymmetric case.
The family of real valued functions will be replaced by a family of complex valued functions, allowing for 
the angular moment parts. It will also be possible to collect all moments into a complex valued function $y$, 
where the moments appear as the derivatives of $y$ at $0$. Again, the complex part of
$y$ is related to the angular moment parts of the multipole moments. The derivation in 
\cite{herb} used a two-surface $S$, which reflected the axisymmetry of the spacetime. In this paper
we will produce the scalar recursion directly, with the methods presented in section \ref{momentsscalar}.

As an application, we will calculate the multipole moments of the Kerr solution.
Another issue is the choice of potential. 
We will show that the potential used by Hansen and the Ernst potential \cite{ernst}, \cite{fodor}
(as well as a large class of other 
potentials) give the same multipole-moments.

\section{Multipole moments of stationary spacetimes} \label{tmoments}
In this section we quote the definition given by Hansen in \cite{hansen}.
We thus consider a stationary spacetime $(M, g_{ab})$ with time-like Killing vector field $\xi^a$.
We let $\lambda=-\xi^a\xi_a$ be the norm, and define the twist $\omega$ through 
$\nabla_a\omega=\epsilon_{abcd}\xi^b\nabla^c\xi^d$.
If $V$ is the 3-manifold of trajectories, the metric $g_{ab}$ (with signature $(-,+,+,+)$) induces the 
positive definite metric
$$h_{ab}=\lambda g_{ab}+\xi_a\xi_b$$ on $V$.
It is required that $V$ is asymptotically flat, i.e.,  there exists a 3-manifold
$\widetilde V$ and a conformal factor $\Omega$ satisfying
\begin{itemize}
\item[(i)]{$\widetilde V = V \cup \Lambda$, \; where $\Lambda$ is a single point}
\item[(ii)]{$\tilde h_{ab}=\Omega^2 h_{ab}$ is a smooth metric on $\widetilde V$}
\item[(iii)]{At $\Lambda$, $\Omega=0, \tilde D_a \Omega =0, \tilde D_a \tilde  
D_b \Omega = 2 \tilde h_{ab}$,}
\end{itemize}
where $\tilde D_a$ is the derivative operator associated with $\tilde  
h_{ab}$.
On $M$, and/or $V$ one defines the scalar potential
$$\phi=\phi_M+i\phi_J, \quad \phi_M=\frac{\lambda^2+\omega^2-1}{4\lambda}, \,\phi_J=\frac{\omega}{2\lambda}$$
The multipole moments of $M$ are then defined on $\widetilde V$ as certain  
derivatives of the scalar
potential $\tilde \phi=\phi/\sqrt \Omega$ at $\Lambda$. More  
explicitly, following \cite{hansen}, let $\widetilde R_{ab}$ denote
the Ricci tensor of $\widetilde V$, and let $P=\tilde \phi$. Define the  
sequence $P, P_{a_1}, P_{a_1a_2}, \ldots$
of tensors recursively:
\begin{equation} \label{orgrec} P_{a_1 \ldots a_n}=C[
\tilde D_{a_1}P_{a_2 \ldots a_n}-
\tfrac{(n-1)(2n-3)}{2}\widetilde R_{a_1 a_2}P_{a_3 \ldots a_n}],
\end{equation}
where $C[\ \cdot \ ]$ stands for taking the totally symmetric and  
trace-free part. The multipole moments
of $M$ are then defined as the tensors $P_{a_1 \ldots a_n}$ at  
$\Lambda$. The requirement that all $P_{a_1 \ldots a_n}$ be totally symmetric and
trace-free makes the actual calculations non-trivial. In the axisymmetric case, 
however, we will see that the tensorial recursion can be replaced by a scalar recursion.

\subsection{Multipole moments of axisymmetric spacetimes} \label{tamoments}
If, in addition to the requirement that $M$ is stationary and asymptotically  
flat, we also impose the condition that $M$ is axisymmetric, the metric can 
 be written in the following canonical form \cite{wald} 
\begin{equation} \label{M}
ds^2=-\lambda(dt-Wd\varphi)^2+\lambda^{-1}(R^2d\varphi^2+e^{2\beta}(dR^2+dZ^2)),
\end{equation}
where $\xi^a=(\frac{\partial}{\partial t})^a$ and
$(\frac{\partial}{\partial \varphi})^a$
are the timelike and axial Killing vectors.
This implies that the metric on $V$ is
$$h_{ab}=\lambda g_{ab}+\xi_a\xi_b\sim R^2d\varphi^2+e^{2\beta}(dR^2+dZ^2).$$
To conformally compactify $V$, we define new variables $\tilde\rho$,
$\tilde z$ and $r, \theta$ via
$\tilde\rho=\frac{R}{R^2+Z^2}=r\sin\theta$, 
$\tilde z=\frac{Z}{R^2+Z^2}=r\cos\theta$ and put $\hat\Omega=r^2e^{-\beta}$.
We then get the rescaled metric, i.e., the metric on $\widetilde V$ as
\begin{equation} \label{omskmet}
\hat h_{ab}= \hat\Omega^2 h_{ab}\sim \tilde\rho^2 e^{-2\beta}d\varphi^2+d\tilde\rho^2+d\tilde z^2
= r^2\sin^2\theta e^{-2\beta}d\varphi^2+dr^2+r^2d\theta^2.
\end{equation}
Therefore we can assume that the rescaled metric has the form
\eqref{omskmet}, where the infinity point $\Lambda$ corresponds to the
point $r=0$. 
However other choices of variables and conformal factors may also give
the rescaled metric the form \eqref{omskmet}. Here we only require
that the rescaled metric has the form \eqref{omskmet}, but 
we do not require that the original metric has the form \eqref{M}.

The conformal factor $\hat\Omega$ is not uniquely determined. One can make a further
conformal transformation of $\widetilde V$, using as conformal factor 
$e^{\kappa}$, where
$\kappa$ is any smooth function on $\widetilde V$ with  
$\kappa(\Lambda)=0$. Thus
$\kappa$ reflects the freedom in
choosing $\hat\Omega$. Of particular importance is the
value of $(\nabla_a \kappa)(\Lambda)=\kappa'(0)$. Namely, under a change
$\hat\Omega \to \Omega=\hat\Omega e^\kappa$, a non-zero $\kappa'(0)$ changes
the moments defined by \eqref{orgrec} in a way which
corresponds to a `translation' of the physical space \cite{geroch}.
With this extra conformal factor the metric becomes
\begin{equation} \label{omskmet2}
\tilde h_{ab} \sim e^{2\kappa}(\tilde\rho^2 e^{-2\beta}d\varphi^2+d\tilde\rho^2+d\tilde z^2)
=e^{2\kappa}(r^2\sin^2\theta e^{-2\beta}d\varphi^2+dr^2+r^2d\theta^2).
\end{equation}
By choosing $\kappa'(0)$ such that the expansion is taken around the generalised centre of mass \cite{beigAPA},
the multipole moments are fixed
and invariant under the restricted remaining conformal freedom. Also,
from their very construction,  the multipole moments are coordinate-independent.

\section{Multipole moments through a scalar recursion on $\mathbb{R}^2$}\label{momentsscalar}
In this section, we will show how the assumption of axisymmetry allows us to replace
the tensors in \eqref{orgrec} by family ${f_n}$ of recursively defined
functions on $\mathbb{R}^2$. 
The multipole moments of $M$ will appear as the values  of ${f_n}$ at the origin point.
The reason that this works is that the following lemma holds:

\begin{lemma}\label{etalemma}
Suppose $\widetilde V$ is a 3-manifold with metric given by \eqref{omskmet2}. Then there exists a
regularly direction dependent (at $\Lambda$) vector field $\eta^a$ with the following properties:\\
a) For all tensors $T_{a_1 \ldots a_n}$, $\eta^{a_1} \ldots
\eta^{a_n}T_{a_1 \ldots a_n}=\eta^{a_1} \ldots \eta^{a_n}C[T_{a_1
  \ldots a_n}]$,\\
b) At $\Lambda$, $P_{a_1 \ldots a_n}$ is determined by 
$\eta^{a_1} \ldots \eta^{a_n}P_{a_1 \ldots a_n}$\\
c) $\eta^a \tilde D_a \eta^b$ is parallel to $\eta^b$.
\end{lemma}

\noindent {\bf Proof:}\\
a)
It is sufficient to require that $\eta_a\eta^a=0$, i.e., that $\eta^a$ is a complex
null vector on $\widetilde V$.
To get the totally symmetric and trace-free part of a tensor $T_{a_1\ldots a_n}$
we first symmetrize and define $S_{a_1\ldots a_n}=T_{(a_1\ldots a_n)}$.
We then subtract all traces to get
$C[T_{a_1\ldots a_n}]=S_{a_1\ldots a_n}-\gamma(n)\tilde h_{(a_1a_2}S_{a_3\ldots a_n)}{}^b{}_b
-\delta(n)\tilde h_{(a_1a_2}h_{a_3 a_4}S_{a_5\ldots a_n)}{}^b{}_b{}^c{}_c-\ldots$,
where $\gamma(n), \delta(n), \ldots$ have their appropriate values and where all terms except 
the first contain $\tilde h_{a_ia_j}$ for some $i,j$. From $\eta_a\eta^a=0$ it follows that
$\eta^{a_1} \ldots \eta^{a_n}C[T_{a_1\ldots a_n}]
=\eta^{a_1} \ldots \eta^{a_n}S_{a_1\ldots a_n}=\eta^{a_1} \ldots \eta^{a_n}T_{a_1\ldots a_n}$,
where the last equality follows from the fact that $\eta^{a_1} \ldots \eta^{a_n}$ is totally
symmetric.\\
b) 
Define 
\begin{equation} \label{etadef}
\eta^a=(\frac{\partial}{\partial\tilde z})^a-i(\frac{\partial}{\partial\tilde\rho})^a
=e^{-i\theta}\left ( (\frac{\partial}{\partial r})^a
-\frac{i}{r}(\frac{\partial}{\partial \theta})^a\right ).
\end{equation}
We see that $\eta_a\eta^a=0$ so a) is valid.
The axisymmetry implies \cite{geroch} that at $\Lambda$,  
$P_{a_1a_2\dots a_n}$ is proportional to $C[z_{a_1}z_{a_2}\dots z_{a_n}]$, i.e.,
\begin{equation}
P_{a_1a_2\dots a_n}(\Lambda)=m_n C[z_{a_1}z_{a_2}\dots z_{a_n}],
\end{equation}
where $z_a=(d\tilde z)_a$ is the direction along the symmetry axis.
Hence, at $\Lambda$,
\begin{equation}\label{etaP}
\eta^{a_1} \ldots \eta^{a_n}P_{a_1 \ldots a_n}
=m_n\eta^{a_1}\dots\eta^{a_n}z_{a_1}\dots z_{a_n}=m_n.
\end{equation} 
Therefore, at $\Lambda$, $P_{a_1 \ldots a_n}$ is determined by 
$\eta^{a_1} \ldots \eta^{a_n}P_{a_1 \ldots a_n}$.\\
c)
With $\eta^a$ defined by \eqref{etadef}, and using the metric (\ref{omskmet2}),
 a direct calculation gives 
\begin{equation} \label{etadeleta}
\eta^a\tilde D_a\eta^b=2\eta^b\eta^c\tilde D_c\kappa.
\end{equation}

Note that the special case $\kappa=0$, i.e., when the metric is of the
form \eqref{omskmet}, gives $\eta^a\tilde D_a\eta^b=\eta^a\hat D_a\eta^b=0$. It is also
important to note that when we interpret $\kappa, \beta, \tilde \phi$ as functions of
two variables, say, $\tilde z, \tilde \rho$, that they are defined in the half-space $\tilde \rho \geq 0$,
but that they can be naturally extended to all $\tilde\rho$ via $\beta(\tilde z, -\tilde \rho)=
\beta(\tilde z,\tilde \rho)$, etc., so that they are even in $\tilde\rho$.  Moreover, $\eta^a$ is direction
dependent at $\Lambda$ when regarded as a vector field on $\widetilde V$, but not when
regarded as a vector field on the half-space $\tilde \rho \geq 0$. Moreover, $\eta^a$ can be naturally
and smoothly extended to $\tilde \rho<0$, and in particular, derivatives
like $\eta^a \tilde \nabla_a \beta$ will then be smooth at $\Lambda$ but not necessarily even
when extended to all values of $\tilde \rho$.

We are now ready to simplify the recursion \eqref{orgrec}. We start by defining
\begin{equation} \label{fn}
f_n=\eta^{a_1}\eta^{a_2}\dots\eta^{a_n}P_{a_1a_2\dots a_n}, \ n=0,1,2,\ \ldots
\end{equation}
In particular, $f_0=P=\tilde\phi=\Omega^{-\frac{1}{2}}\phi$. By contracting \eqref{orgrec}
with $\eta^a$ we get the following theorem.
\begin{theorem}\label{scalrarrec2d}
Let $\eta^a$ have the properties given by lemma \ref{etalemma}, and let $f_n$
be defined by \eqref{fn}. We then have the recursion
\begin{equation}\label{scalarrec}
f_n=\eta^a\tilde D_af_{n-1}-2(n-1)f_{n-1}\eta^a\tilde D_a\kappa
-\tfrac{(n-1)(2n-3)}{2}\eta^a\eta^b\widetilde R_{ab}f_{n-2},
\end{equation}
where the moments $m_n$ are given by $m_n=f_n(\Lambda)$.
\end{theorem}
{\bf Proof:}\\
Let
\begin{equation}
 T_{a_1 \ldots a_n}=\tilde D_{a_1}P_{a_2 \ldots a_n}-
\tfrac{(n-1)(2n-3)}{2}\widetilde R_{a_1 a_2}P_{a_3 \ldots a_n}.
\end{equation}
Using the recursion \eqref{orgrec} and the properties of $\eta^a$ we get
\begin{align*}
f_n&=\eta^{a_1}\dots\eta^{a_n}C[T_{a_1\dots a_n}]
=\eta^{a_1}\dots\eta^{a_n}T_{a_1\dots a_n}\\
&=\eta^{a_2}\dots\eta^{a_n}\eta^{a_1}\tilde D_{a_1}P_{a_2 \ldots a_n}-
\tfrac{(n-1)(2n-3)}{2}\eta^{a_1}\eta^{a_2}\widetilde R_{a_1 a_2}\eta^{a_3}\dots\eta^{a_n}P_{a_3 \ldots a_n}\\
&=\eta^{a}\tilde D_{a}f_{n-1}-P_{a_2\dots a_n}\eta^a\tilde D_{a}(\eta^{a_2}\dots\eta^{a_n})
-\tfrac{(n-1)(2n-3)}{2}\eta^{a}\eta^{b}\widetilde R_{ab}f_{n-2}.
\end{align*}
Using  \eqref{etadeleta} we get \eqref{scalarrec}. By comparing \eqref{etaP} and \eqref{fn},
we see that $m_n=f_n(\Lambda)$.

Note that the proof was carried out in $\widetilde V$, but the resulting scalar recursion
is most conveniently regarded as defined on $\mathbb{R}^2$ with Cartesian coordinates
$\tilde z, \tilde \rho$ or polar coordinates $r,\theta$. In $\eta^a\eta^b\widetilde R_{ab}$,
$\widetilde R_{ab}$ refers to the Ricci tensor of $\widetilde V$; using the metric
\eqref{omskmet2} one readily finds
\begin{equation}\label{etaRab}
\eta^a\eta^b\widetilde R_{ab}
=\eta^a \tilde D_a(\eta^b \tilde D_b \beta)-(\eta^a\tilde D_a \beta)^2-\frac{2i}{\tilde \rho}\eta^a\tilde D_a\beta
-\eta^a \tilde D_a(\eta^b \tilde D_b \kappa)+(\eta^a\tilde D_a \kappa)^2.
\end{equation}
{\bf Remark:}\\
As well as different sign conventions some authors use the 
convention $M_n=\frac{1}{n!}\tilde z^{a_1}\dots\tilde z^{a_n}P_{a_1\dots a_n}(\Lambda)$.
This implies that $M_n=\pm \frac{2^nn!}{(2n)!}m_n$.
\section{Multipole moments through a scalar recursion on $\mathbb{R}$}
In this section we show that we can reduce the problem from scalar fields of two variables to scalar fields of one variable.
So far we have not required analyticity of the fields. 
There are several proofs of analyticity for the potential and the metric
after a special rescaling in the case with non-zero mass \cite{beigPRSL}  \cite{kundu}.
However, since we don't want to exclude spacetimes with zero mass, analyticity will be assumed explicitly.
In this article, analyticity will always be taken in the 'real' sense, i.e. that the
 fields are given by power series with positive radii of convergence. 
In the cases where the results of this paper are used to compute the multipole moments for a certain spacetime, it is easy
to check analyticity.

\begin{lemma}\label{analyticitylemmma}
Let $\beta$  come from (\ref{omskmet}), $\tilde\phi$ be the rescaled (axisymmetric) potential,
and
suppose that $\beta$ and $\tilde\phi$  are analytic in a neighbourhood of $\Lambda \in \tilde V$.
Then $\beta(\tilde z,\tilde\rho)$ contains a factor $\tilde\rho^2$.
Furthermore, viewed as a functions on $\mathbb{R}^2$, $\eta^a\eta^b\widetilde R_{ab}$ and $f_n$ are analytic in $(\tilde z,\tilde\rho)$ in a neighbourhood of $(0,0)$.
\end{lemma}
{\bf Proof:}\\
First we establish that $\beta$ contains a factor $\tilde\rho^2$.
Consider an orbit $\Gamma:\{\tilde \rho=\rho_0\}$ in the surface $\tilde z=z_0$. The circumference 
$C(z_0,\rho_0)=\int_\Gamma{ds}=\int_0^{2\pi}{\rho_0 e^{-\beta(z_0,\rho_0)}d\phi}=2\pi\rho_0 e^{-\beta(z_0,\rho_0)}$.
We also know \cite{bianchi} that $C(z_0,\rho_0)=2\pi\rho_0-\frac{\pi K_0\rho_0^3}{3}+\mathcal{O}(\rho_0^4)$, where $K_0$ is the scalar curvature at $\tilde \rho=0$.
Hence
\begin{align*}
\frac{\pi K_0}{3}&=\lim_{\rho_0\rightarrow 0^+}{\frac{2\pi\rho_0-C(z_0,\rho_0)}{\rho_0^3}}\\
&=\lim_{\rho_0\rightarrow 0^+}{\frac{2\pi}{\rho_0^2}(1-e^{-\beta(z_0,\rho_0)})}
=\lim_{\rho_0\rightarrow 0^+}{\frac{2\pi\beta(z_0,\rho_0)}{\rho_0^2}(1+\mathcal{O}(\beta(z_0,\rho_0)))}.
\end{align*} 
Since $\tilde z=z_0$ is a smooth manifold, $K_0$ is finite, and we have that $\beta(\tilde z,\tilde \rho)$ 
contains the factor $\tilde \rho^2=r^2\sin^2\theta$.
{}From this we see that $\frac{2i}{\tilde \rho}\eta^a\tilde D_a\beta$ is analytic in $(\tilde z,\tilde\rho)$
and consequently all terms in \eqref{etaRab} are analytic in  $(\tilde z,\tilde\rho)$. 
The analyticity of $f_0=\tilde\phi$ is assumed. It follows inductively from \eqref{scalarrec} 
that all $f_n$ are analytic in $(\tilde z,\tilde\rho)$. 

To obtain a scalar recursion on $\mathbb{R}$, we must extract/define 
functions of one real variable from our given functions on $\mathbb{R}^2$, where the latter are supposed to be analytic
in terms of $(\tilde z,\tilde\rho)$. 

\begin{definition} \label{ldef}
Suppose that $g:\mathbb{R}^2\rightarrow \mathbb{C}$ is an analytic function 
of two variables in a neighbourhood of the origin. We then define the leading order
function  $g_L$ on $\mathbb{R}$ via $g_L(r)=g(r,-i r)$.
\end{definition}
Note that the analyticity condition allows us to write
$g(\tilde z,\tilde\rho)=\sum_0^\infty \sum_0^\infty a_{nk}\tilde z^n \tilde\rho^k$ for 
$(\tilde z,\tilde\rho)$ in some disc $B_{\delta}(0)$ centred at the origin. Thus $\tilde z,\tilde\rho$
can be allowed to be complex in $B_{\delta}(0)$ and the definition $g_L(r)=g(r,-i r)$
makes sense as long as $r\sqrt{2}<\delta$.

We will now fix $\eta^a$ to be precisely 
$\eta^a=(\frac{\partial}{\partial\tilde z})^a-i(\frac{\partial}{\partial\tilde\rho})^a$.
That the recursion \eqref{fn} can be replaced by a recursion of the leading order functions,
will then follow from the following lemma:

\begin{lemma} \label{derlemma}
If $g$ satisfies the conditions of definition \ref{ldef}, $(\eta^a \tilde D_a g)_L(r)=g_L'(r)$.
\end{lemma}
{\bf Proof:}
Both sides equal $\frac{\partial g}{\partial\tilde z}(r,-ir)-i\frac{\partial g}{\partial\tilde\rho}(r,-ir)$.

We are now ready to derive the promised scalar recursion on $\mathbb{R}$.

\begin{theorem} \label{recmoments}
Suppose that the conditions in lemma \ref{analyticitylemmma} are met,  and
define $y_n(r)=(f_n(\tilde z,\tilde\rho))_L(r)$. Then the multipole moments $m_n$ are given
through the recursion 
\begin{equation}\label{realrec}
y_n=y_{n-1}'-2(n-1)\kappa_L'y_{n-1}-\tfrac{(n-1)(2n-3)}{2}My_{n-2},
\end{equation}
where 
\begin{equation} \label{ml}
M(r)=\beta_L''-(\beta_L')^2+\frac{2}{r}\beta_L'-\kappa_L''+(\kappa_L')^2
\end{equation}
and $m_n=y_n(0)$.
\end{theorem}

\noindent {\bf Proof:}
Taking the leading order part of \eqref{scalarrec}, putting
$M(r)=(\eta^a\eta^b\widetilde R_{ab})_L(r)$, and applying lemma \ref{derlemma},
we get \eqref{realrec}. Also,  $m_n=f_n(0,0)=y_n(0)$.

\section{All moments from one scalar function}
The recursion \eqref{realrec} would simplify considerably if $M=0$, since we then get a recursion
of depth one. As is seen below, it is possible to 
choose $\kappa_L$ such that $M=0$. The simplified recursion is then solved by choosing a different 
radial parameter $\rho$. The choice of $\kappa_L$ in order for $M=0$ to hold in (\ref{realrec}) can be viewed in two ways. 
Either, one starts with the spacetime cast in the form (\ref{omskmet2}) and chose
a particular $\kappa$, or else one uses the metric 
$\hat h_{ab}$ in the form (\ref{omskmet}) as the starting point.
In the latter case, starting with $\hat h_{ab}$, we have the potential $\hat \phi=\phi/\sqrt{\Omega}$,
and perform a second conformal rescaling $\Omega \to \Omega e^{\kappa}$ where $\kappa$
is chosen such that $\kappa_L$ solves equation (\ref{ml}) with $M=0$. The new potential
$\tilde \phi$ is then related to $\hat \phi$ via $\tilde \phi=e^{-\kappa/2} \hat \phi$. 
Although equivalent, the latter viewpoint can be advantageous when explicit
spacetimes are considered (cf. section \ref{kerrsec}).

\begin{theorem}\label{scalarfunc}
Suppose that $M$ is a stationary axisymmetric asymptotically flat spacetime, 
with $\tilde V$ a conformal rescaling of the manifold of time-like Killing-trajectories,
where the metric $\hat h_{ab}$ on $\tilde V$ has the form \eqref{omskmet}. 
Let $\hat\phi$ be the conformally rescaled potential.
Furthermore, assume that $\hat\phi$ and $\beta$ are analytic 
in a neighbourhood of $\Lambda \in \tilde V$.
Choose $\kappa$ such that, for some constant $C$,
\begin{equation} \label{kappal}
\kappa_L(r)=-\ln(1-r\int_0^r{\frac{e^{2\beta_L(r)}-1}{r^2}dr}-r C)+\beta_L(r),
\end{equation}
thereby inducing the metric $\tilde h_{ab}=e^{2\kappa}\hat h_{ab}$.
Define $r(\rho)$ implicitly by $\rho(r)=re^{\kappa_L-\beta_L}$.
Put $y(\rho)=\tilde \phi_L(r(\rho))=e^{-\kappa_L(r(\rho))/2}\hat\phi_L(r(\rho))$. Then the multipole moments
$m_0,m_1,\dots$ of $M$ are given by $m_n=\frac{d^ny}{d\rho^n}(0)$.
\end{theorem}
{\bf Remark:}\\
In the expression for $\kappa_L$, the constant $C$ is seen to equal $\kappa_L'(0)=\kappa'(\Lambda)$.
In particular one can choose $\kappa_L'(0)$ such that the expansion is taken around the generalised centre of mass.
With this choice the multipole moments are unique.
This is accomplished in the following way: 
If $k$ is the first integer such that $m_k\neq 0$, we choose $\kappa_L'(0)$ so that $m_{k+1}=0$.
This is equivalent to the definition \cite{beigAPA} of generalised centre of mass:
$0=P_{a_1\dots a_{k+1}}(\Lambda)P^{a_1\dots a_k}(\Lambda)=m_{k+1}m_{k}C[z_{a_1}\dots z_{a_{k+1}}]C[z^{a_1}\dots z^{a_k}]
\propto m_{k+1}m_k z_{a_{k+1}}$.
Note also that the non-leading terms of $\kappa$ can be chosen arbitrarily.

\noindent {\bf Proof:}\\
{}From lemma \ref{analyticitylemmma}, it follows that $\beta_L(0)=\beta_L'(0)=0$.
Taking this into account, a direct insertion of \eqref{kappal} into \eqref{ml} shows that
$M(r)=0$, i.e., this choice of $\kappa_L$ solves the differential equation $M(r)=0$.
Therefore, with this choice of $\kappa_L$ the recursion \eqref{realrec} reduces to
\begin{equation} \label{ynrec}
y_n=y_{n-1}'-2(n-1)\kappa_L'y_{n-1}.
\end{equation}
Put $z_n(r)=e^{-2n\kappa_L(r)}y_n(r)$ so that $z_n(0)=y_n(0)=m_n$.
 As in \cite{herb}, \eqref{ynrec} becomes
\begin{equation} \label{znrec}
z_n(r)=e^{-2\kappa_L(r)}z'_{n-1}(r)
\end{equation}
If $\rho$ is such that
$
\frac{d\rho(r)}{dr}=e^{2\kappa_L(r)},
$
we find that $z_n=\frac{dz_{n-1}}{d\rho}=\frac{d^nz_0}{d\rho^n}$.
The specific choice \eqref{kappal} of $\kappa_L$ implies that $r(\kappa_L'-\beta_L')+1=e^{\kappa_L+\beta_L}$.
Hence,
\begin{equation}\label{rhodef}
\rho(r)=\int_0^r{e^{2\kappa_L}dr}
=\int_0^r{(r(\kappa_L'-\beta_L')+1)e^{\kappa_L-\beta_L}dr}
=re^{\kappa_L-\beta_L}.
\end{equation}
Thus equation \eqref{rhodef} implicitly defines $r(\rho)$ in a neighbourhood of $\rho=0$.
We can now define
$y(\rho)=z_0(r(\rho))=\tilde\phi_L(r(\rho))$
and get all the moments from the function $y(\rho)$:
\begin{equation}
m_n=\frac{d^ny}{d\rho^n}(0).
\end{equation}

\section{The Kerr solution} \label{kerrsec}
As an example of how our method can be used, we compute the multipole moments for the Kerr solution.
Here we will not use the Weyl canonical coordinates because the expressions would be much longer than necessary.
Instead we will follow Hansen \cite{hansen}. 
We begin with Boyer-Lindquist coordinates $(\tilde r, \theta,\varphi)$ and the metric
\begin{equation}
ds^2=\frac{\tilde r^2-2m\tilde r+a^2\cos^2\theta}{\tilde r^2-2m\tilde r+a^2}d\tilde r^2+(\tilde r^2-2m\tilde r+a^2\cos^2\theta)d\theta^2+(\tilde r^2-2m\tilde r+a^2)d\varphi^2.
\end{equation}
Following Hansen we define a new radial coordinate $r$
through $\tilde r=r^{-1}(1+mr+\frac{1}{4}(m^2-a^2)r^2)$ and use the conformal factor
$\hat \Omega=\frac{r^2}{\sqrt{(1-\frac{1}{4}(m^2-a^2)r^2)^2-a^2r^2\sin^2\theta}}$.
Define $\tilde z=r\cos\theta$ and  $\tilde\rho=r\sin\theta$.
We then get
\begin{align}
\beta(\tilde z,\tilde\rho)&=\frac{1}{2}\ln\left(1-\frac{(4a\tilde\rho)^2}{(4-(m^2-a^2)(\tilde z^2+\tilde\rho^2))^2}\right),\\
\beta_L(r)&=\frac{1}{2}\ln(1+a^2r^2),\\
\hat\phi(\tilde z,\tilde\rho)&=\frac{m(1+\frac{1}{4}(m^2-a^2)(\tilde z^2+\tilde\rho^2)-ia\tilde z)}
{((1-\frac{1}{4}(m^2-a^2)(\tilde z^2+\tilde\rho^2)^2)^2-a^2\tilde\rho^2)^{\frac{3}{4}}},\\
\hat\phi_L(r)&=\frac{m(1-iar)}{(1+a^2r^2)^{3/4}}.
\end{align}
Furthermore, from theorem \ref{scalarfunc} we get
\begin{align}
\kappa_L(r)&=-\frac{1}{2}\ln\left(\frac{(r^2a^2+r\kappa'(0)-1)^2}{r^2a^2+1}\right),\\
\rho(r)&=\frac{r}{1-r\kappa'(0)-r^2a^2},\\
r(\rho)&=\frac{\sqrt{(\rho\kappa'(0)+1)^2+4a^2\rho^2}-\rho\kappa'(0)-1}{2\rho a^2},\\
y(\rho)&=e^{-\kappa_L(r)/2}\hat\phi_L(r)=\frac{m\sqrt{1-r\kappa'(0)-a^2r^2}}{1+iar}=\frac{m}{\sqrt{1+(2ia+\kappa'(0))\rho}}.
\end{align}
Comparing with the Schwarzschild solution, $a=0$, we see that the Kerr solution is in a sense 
 an imaginary translation of the Schwarzschild spacetime.
Similar descriptions can be found in \cite{schiffer} and \cite{newman}.

With $\kappa'(0)=0$, we get the expansion around the centre of mass:
\begin{equation}
\sum_{n=0}^\infty{\frac{m_n\rho^n}{n!}}=y(\rho)=
\sum_{n=0}^\infty{\frac{m(2ia\rho)^n\sqrt{\pi}}{n!\Gamma(\frac{1}{2}-n)}}.
\end{equation}
With the conventions used by Hansen \cite{hansen} we have
\begin{equation}
M_n=\frac{-2^nn!}{(2n)!}m_n=\frac{-2^nn!m(2ia)^n\sqrt{\pi}}{(2n)!\Gamma(\frac{1}{2}-n)}=-(-i)^{n}ma^n.
\end{equation}
Thus our calculation agrees with Hansen \cite{hansen}. (In \cite{hansen}, the moments were
given without a proof, see however \cite{simon} \cite{sotir}.)

\section{Potentials}\label{potsec}
So far we have only used the complex version of the potential defined by Hansen \cite{hansen}
\footnote{We have changed sign compared to \cite{hansen}}
\begin{equation} \label{Hpot}
\phi_H=-\frac{\lambda^2-1+\omega^2+2i\omega}{4\lambda},
\end{equation}
and as Hansen points out, the choice of potential is not unique. 
This freedom of choice was addressed 
in \cite{simonJMP} using a different but equivalent definition of multipole moments compared to the 
Geroch-Hansen definition. Here we will however only use theorem \ref{recmoments} to investigate the freedom
in the choice of potential.

For instance, in the special case
when the stationary axisymmetric spacetime is actually static, the Hansen potential reduces to
\begin{equation} \label{hpot}
\phi_H=\frac{1-\lambda^2}{4\lambda}.
\end{equation}
This potential is not the same as the one given by Geroch, who uses\footnote{Again, we have changed
sign. The factor $\lambda^{-1/4}$ comes from different definitions of the metric for $V$.}
\begin{equation}\label{gpot}
\phi_G=\lambda^{-1/4}(1-\sqrt\lambda).
\end{equation}
It is not obvious that \eqref{hpot} and \eqref{gpot} give the same sequence of multipole 
moments. This was however shown in \cite{simonJMP}. 
The following theorem proves the same thing in the axisymmetric case using a different technique.
\begin{theorem}
Suppose that the spacetime $M$ is static and axisymmetric. 
Then the potentials
\eqref{hpot} and \eqref{gpot} produce the same moments.
\end{theorem}
\noindent {\bf Proof:}\\
We may assume that the coordinates are chosen in the way described in section \ref{tamoments} with 
$\kappa=0$, i.e., $\tilde\Omega=r^2e^{-\beta}=(\tilde \rho^2+\tilde z^2)e^{-\beta}$;
in other words we use Weyl coordinates.
With this choice of coordinates, $\tilde\alpha=-\frac{\ln\lambda}{2r}$ is (flat-) harmonic
with respect to $(\tilde z,\tilde\rho,\varphi)$ taken as cylindrical coordinates in $\mathbb{R}^3$.
{}From this it follows that $\frac{\phi_G}{r}=\frac{2}{r}\sinh{\frac{\tilde\alpha r}{2}}$ is analytic
near $\Lambda \in \tilde V$, since $\frac{2}{r}\sinh{\frac{\tilde\alpha r}{2}}$ is even in $r$. It also follows 
from the explicit form of $\beta$ in \cite{herb}
that $\beta$ is analytic
and thus $\Omega$ is analytic with $\Omega_L=0$. 
{}From the definitions \eqref{hpot} and \eqref{gpot} we see that
\begin{equation}\label{gpottohpot}
\tilde\phi_H=\frac{\tilde\phi_G}{4}(\Omega\tilde\phi_G^2+2)\sqrt{\Omega\tilde\phi_G^2+4},
\end{equation}
so that $\tilde \phi_H$
is also analytic. Taking leading order functions on both sides implies
$(\tilde \phi_H)_L=(\tilde \phi_G)_L$.
Referring to theorem \ref{recmoments}, this is sufficient for both potentials to generate the same multipole moments.

Now consider the stationary case again. 
Assume that $\tilde\phi_H$  and $\Omega$ are analytic functions in a neighbourhood of $\Lambda$. 
The same will then hold for $\tilde\phi_M$ and $\tilde\phi_J$, the real
and imaginary part of $\tilde\phi_H$ respectively. Asymptotic flatness implies that $\Omega_L=0$.
Using theorem \ref{recmoments} we see that we can in fact choose any well behaved potential that satisfies 
$(\tilde\phi)_L=(\tilde\phi_H)_L$ to get the same moments.
For instance if we wanted to use the Ernst potential \cite{ernst}, \cite{fodor} instead,
\begin{equation} \label{epot}
\phi_E=\frac{1-\lambda-i\omega}{1+\lambda+i\omega},
\end{equation}
we see that
\begin{equation}
\tilde\phi_E=\frac{2\tilde\phi_H}{1+\sqrt{4\Omega\tilde\phi_M^2+4\Omega\tilde\phi_J^2+1}}.
\end{equation}
Taking leading order functions of both sides, we find that $(\tilde\phi_E)_L=(\tilde\phi_H)_L$,
and thus the potentials produce the same moments. This proves the following theorem.
\begin{theorem}
Suppose that the spacetime $M$ is stationary and axisymmetric. 
Then the potentials
\eqref{Hpot} and \eqref{epot} produce the same moments.
\end{theorem}
{\bf Remark}\\
This result could also be obtained easily by the method in \cite{simonJMP}.
\section{Discussion}
We have shown that the relatively complicated tensorial recursion \eqref{orgrec} of Geroch and Hansen can, 
in the stationary axisymmetric case, 
be reduced to a recursion of scalar functions.
Furthermore, we have demonstrated how a careful choice of conformal factor collects all 
moments into one complex valued function on $\mathbb{R}$, where the moments appear as the derivatives at $0$. 
These results reduce to the results in \cite{herb} in the static case. The main ideas in \cite{herb} and in this paper
are similar, but here we have simplified the proofs as well as generalised the results.
As an application of this method, we have shown how easily the moments of the Kerr solution follow. 
We have also seen that there is a great freedom in the choice of the potential. This can possibly be used to simplify
the field equations expressed in terms of the potential. 

In \cite{backdahl} it was shown how to obtain the metric for a static axisymmetric spacetime with prescribed 
multipole moments. With the methods presented here it is natural to try to extend these results to the stationary case.
Such results would be very desirable due to the physical relevance of the multipole moments. For instance, 
axisymmetric stationary solutions are good approximations
for astrophysical objects. 

It would also be interesting to address the growth condition of the multipole moments
for existence of stationary spacetimes. 
Another natural extension of the work presented here is of course the
 general non-axisymmetric case.

\end{document}